\pdfoutput=1
\documentclass[aps,pre,twocolumn,groupedaddress,longbibliography]{revtex4-1}
\usepackage[utf8]{inputenc}

\usepackage{color}
\usepackage{amsmath}
\usepackage{amssymb}
\usepackage{bm}
\usepackage{graphicx}
\usepackage{hyperref}

\usepackage{array}

\definecolor{bleu}{rgb}{0,0,0}

\DeclareMathOperator{\sign}{sgn}

\begin{document}

\title{Topological elasticity of non-orientable ribbons}
\author{Denis Bartolo and David Carpentier}
\affiliation{Univ. Lyon, ENS de Lyon, Univ. Claude Bernard, CNRS, Laboratoire de Physique, F-69342, Lyon.}

\date{\today}

\begin{abstract}
In this article, we unravel an intimate relationship between  two seemingly unrelated concepts: elasticity, that defines the local relations between stress and strain of deformable bodies, and  topology that classifies their global shape. 
Focusing on M\"obius strips, we establish that  the 
elastic response of surfaces with  non-orientable topology is: non-additive, non-reciprocal and contingent on stress-history. Investigating the  elastic instabilities of non-orientable ribbons, we then challenge the very concept of bulk-boundary-correspondence of topological phases. We   establish  a quantitative connection between the  modes found at the interface between inequivalent topological  insulators and solitonic bending excitations  that freely propagate through the bulk non-orientable ribbons. 
 Beyond the specifics of mechanics, we argue that non-orientability offers a versatile platform to tailor the response of systems as diverse as  liquid crystals, photonic and electronic matter.

\end{abstract}

\maketitle

\section{Introduction}
Sewing the first piece of fabric, prehistoric men  laid out  the first  principles of  metamaterial design~\cite{Kvavadze2009}:   elementary units assembled into geometrical patterns  form structures with mechanical properties that can surpass those of their constituents~\cite{Bertoldi2017}. 
In the early 2010's, building on quantitative analogies with the topological phases of quantum matter, researchers laid out robust design rules for metamaterials supporting mechanical deformations immune from geometrical and material imperfections
~\cite{KaneLubensky,ProdanProdan,Huber2015,Nash2015,Huber2016,Bertoldi2017,Mao2018}. 
Today, mechanical analogs  of virtually all topological phases of electronic matter have been experimentally realized, or theoretically designed, with  mechanical components as simple as  coupled gyroscopes or lego pegs~\cite{Nash2015,Huber2015,Chen2014,Bertoldi2017,Prodan2018,Susstrunk2016,Mitchell2018}. %
{\color{bleu}The basic strategy 
consists in connecting mechanical systems with  gapped vibrational spectra having topologically distinct eigenspaces~\cite{Hasan2010,Bernevig,Mao2018}. At the interface, this mismatch  causes  a local gap closing revealed by linear edge modes  topologically protected from disorder and backscattering.
Until now, as topological mechanics was  inspired by analogies with condensed matter, it has  been essentially restrained to 
 metamaterials  assembled from repeated mechanical units, that inherit robustness from the topology of their abstract vibrational eigenspace~\cite{Mao2018,Sun2019}.} 

In this article, we elucidate the consequences of real-space topology on the mechanics  of homogeneous materials.
Firstly, we demonstrate that non-orientability  makes M\"obius strips' elasticity: non-additive, nonreciprocal and multistable.
In particular, we  demonstrate how the static deformations of non-orientable surfaces encode their stress history: M\"obius strips have a mechanical memory. {\color{bleu}Secondly, we address the impact of non-orientability on the paradigmatic Euler elastic instability. We show that the associated buckling patterns propagate as solitary waves on M\"obius strips.} We finally establish the equivalence between these  non-linear bulk excitations and the edge modes found at the interface between inequivalent topological states in one-dimensional topological insulators~\cite{Hasan2010}.
\begin{figure}[t!]
\includegraphics[width=\columnwidth,angle=0]{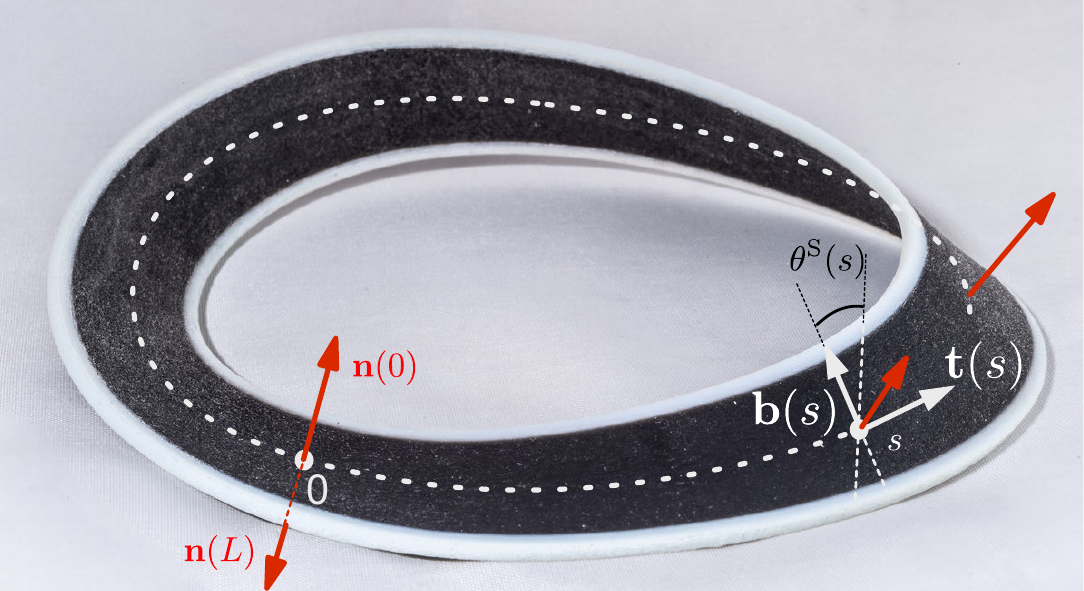}
\caption{{\bf A non-orientable elastic ribbon}. Example of a 3D printed M\"obius strip of width $1\,\rm cm$. 
The  dotted white line indicates the base circle $\mathbf C(s)$. The shear angle $\theta^{\rm s}$ is defined with respect to the local normal to the surface $\mathbf n(s)$ (red arrows). 
Note that $\mathbf n(s)$ reverses its sign after one full turn around the strip $\mathbf n(0)=-\mathbf n(L)$. 
The tangent, $\mathbf t(s)$, and  bi-normal, $\mathbf b(s)$, vectors are indicated with white arrows. }
\label{Fig1}
\end{figure}

\section{Topological elasticity of non-orientable surfaces}
\label{sectionII}
{\color{bleu}Simply put, a non-orientable surface  is a one-sided thin sheet. A paradigmatic example  is given by the M\"obius  strip shown in Fig.~\ref{Fig1} that can be easily replicated by applying a half twist to a band of paper before glueing its two ends. Orientability is indeed a global (topological) property that can be altered only by cutting and gluing back a geometrical surface.} In contrast, linear elasticity  describes  local deformations in response to gentle mechanical stresses. {Before introducing a technical  framework to relate these two seemingly unrelated concepts, let first us gain some intuition about  their relationship.} We consider  the simple example of  a M\"obius strip made of an elastic material showed in Fig.~\ref{Fig1}. The shear deformations of the strip is locally quantified by the angle $\theta^{\rm S}(s)$, where $s\in[0,L]$ indicates the curvilinear coordinates along the strip centerline. {\color{bleu}$\theta^{\rm S}(s)$ is a rotation angle defined  with respect to  the vector $\mathbf n(s)$ normal to the surface}. A direct consequence of non-orientability is that  no stress distribution can yield  homogeneous shear deformations over a M\"obius strip. As illustrated in Fig.~\ref{Fig1}, when transported around the entire strip, $\mathbf n(s)$ changes sign, thereby implying that $\theta^{\rm S}(0)=-\theta^{\rm S}(L)$, and that the shear angle must vanish at least once along the ribbon.  The impossibility to  assign an unambiguous orientation to the surface constrains the ribbon to remain undeformed at one point whatever the magnitude of the applied stress.
We now  account for this topological protection against shear by describing  the  elasticity of non-orientable ribbons as a $\mathbb Z_2$ gauge theory.
\subsection{Orientability as a $\mathbb Z_2$ gauge charge.}
\label{SectionGauge}
\begin{figure}[t!]
\includegraphics[width=\columnwidth,angle=0]{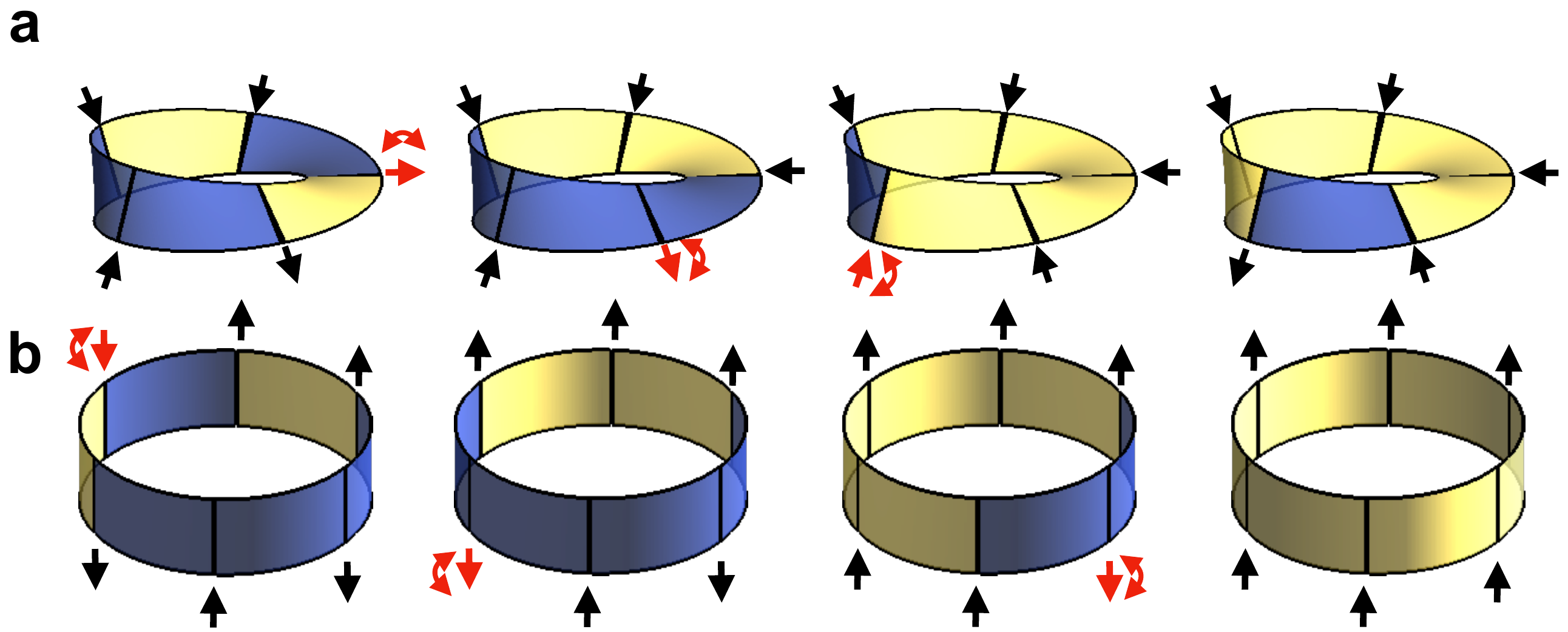}
\caption{{\bf Orientability as a $\mathbb Z_2$ gauge charge}. Two discretized ribbons:  a M{\"o}bius strip ({\bf a.}), and a  cylinder ({\bf b.}). The arrows indicate the orientation of the $\mathbf b_i$ vectors, and the plaquettes' color  the sign of $\eta_{i,i+1}$ on each link, yellow: $\eta_{i,i+1}=+1$, blue: $\eta_{i,i+1}=-1$. 
The gauge transformations described in Section~\ref{SectionGauge} are illustrated by the red arrows.  
{\bf a.} No series of orientation-gauge transformations can result in a homogeneous $\eta$ field on a M\"obis strip: $\mathcal O=-1$.
({\bf b.}) Starting from a heterogeneous $\eta$ field, the gauge transforms  result in a homogeneous $\eta=+1$ field on a orientable cylinder: $\mathcal O=+1$.
}
\label{Fig2}
\end{figure}

For sake of clarity, we  restrain ourselves to strips of constant width $w$ akin to that showed in Figs.~\ref{Fig1}, and~\ref{Fig2}. They are defined as ruled surfaces ${\mathbf S}(s,z)={\mathbf C}(s)+z{\mathbf b}(s)$ where ${\mathbf C}(s)$ is a base circle of perimeter $L$, and  $\mathbf b(s)$ is a unit-vector field  normal to the  tangent-vector field ${\mathbf t}(s)$, see Fig.~\ref{Fig1}. Given this definition $s\in[0,L]$ and $z\in [-w/2,w/2]$. We stress that the direction of $\mathbf b(s)$ is arbitrary: a local transformation $\mathbf b(s)\to \epsilon(s)\mathbf b(s)$, where $\epsilon(s)=\pm1$ leaves the strip geometry unchanged. The tangent to the base circle $\mathbf t(s)$ being unambiguously defined,  the normal vector ${\mathbf n}(s)=\mathbf t(s)\times \mathbf b(s)$ is  defined up to the same $\epsilon(s)$ sign factor as $\mathbf b(s)$.

By definition, non-orientable strips correspond to shapes where the fields $\epsilon(s)\mathbf b(s)$ and $\epsilon(s)\mathbf n(s)$ are  discontinuous regardless of the sign convention $\epsilon(s)$. This intrinsic ambiguity in defining the orientation of the (bi)normal vector is better illustrated when discretizing the strip, see Fig.~\ref{Fig2}. Setting $s=i a$, where $a=L/N$ and $i\in[1,N-1]$, we  introduce the $\mathbb Z_2$ gauge field $\eta_{i,i+1}= \epsilon_i\epsilon_{i+1}$ which represents the connection between adjacent sign conventions.  The topological charge ${\mathcal O}= \prod_{i=1}^{N} \eta_{i,i+1}$, defines the surface orientability: orientable surfaces correspond to $\mathcal O=+1$ and nonorientable ones to $\mathcal O=-1$. The independence of $\mathcal O$ with respect to the  sign convention  becomes clear when applying the series of  gauge transformations sketched in Figs.~\ref{Fig2}a and ~\ref{Fig2}b. Starting from an arbitrary position {\color{bleu}$i_{\rm G}+1$} and moving along the base circle,  wherever a link 
with  $\eta_{i,i+1}=-1$ is found, we change the sign of $\epsilon_{i+1}$. This transformation reverses simultaneously  the signs of both $\eta_{i,i+1}$ and $\eta_{i+1,i+2}$ thereby leaving $\mathcal O$ unchanged. Moving along the strip and
 repeating this procedure, we  find that the  gauge field on all links but the last one can be set to $\eta=+1$. On the last link, it takes the value {\color{bleu}$\eta_{i_{\rm G},i_{\rm G}+1}=\mathcal O$}. Therefore, when $\mathcal O=-1$ there is an obstruction to define a homogeneous surface orientation: the surface is non-orientable.
%
\subsection{Elasticity of twisted elastic strips.}
\label{Section:shear}
We now make use of this geometric framework to describe the elastic response  of a soft M\"obius strip having a stress-free
 equilibrium shape defined by the triad $(\mathbf{t}^0(s),\mathbf{b}^0(s),\mathbf{n}^0(s))$. 
For sake of simplicity, we do not resort to the full Foppl-von Karman theory of elastic plates~\cite{Audoly_Book}. Instead, we consider simplified models to single out 
the impact of non-orientability on shear, twist and bend deformations {\color{black}leaving a more realistic mechanical description for future work}.  
The amplitude of the pure-shear, $\theta^{\rm S}(s)$, and pure-twist angles, $\theta^{\rm T}(s)$, 
are usually defined from the deformation vector 
$\mathbf u(s)=\mathbf b(s)-\mathbf b^0(s) \equiv\theta^{\rm S}(s)\mathbf t^0(s)+\theta^{\rm T}(s)\mathbf n^0(s)$. 
As discussed in the previous section, however, both $\mathbf b(s)$ and $\mathbf n(s)$ are  defined up to a sign
convention $\epsilon(s)$, while all physical quantities must be independent of this arbitrary  choice. We therefore introduce the orientation-independent deformation field: 
\begin{equation}
\left(\epsilon\mathbf u\right)=(\epsilon \theta^{\rm S}) \mathbf t^0 +\theta^{\rm T}(\epsilon\mathbf n^0).
\label{twistshear2}
\end{equation} 
$\left(\epsilon\mathbf u\right)$ is  invariant upon the orientation transformation: 
$\lbrace \epsilon(s)\to-\epsilon(s),\,\theta^{\rm S}(s)\to-\theta^{\rm S}(s),\,
\theta^{\rm T}(s)\to\theta^{\rm T}(s)\rbrace$. 
Due to the possibly  discontinuous nature of the $\epsilon$ field, we first define the harmonic elasticity associated to  $(\epsilon u)$ by resorting to a discretization of the ribbon geometry. 
The simplest harmonic elasticity is then given by 
${\mathcal E}=K/(2a)\sum_i \left[(\epsilon\mathbf u)_{i+1}-(\epsilon\mathbf u)_{i}\right]^2$ where $K$ is an isotropic elastic constant, and $\mathcal E$ is readily recast into: 
\begin{equation}
{\mathcal E}
=
\frac{K}{2a}\sum_i \left[\mathbf u_{i+1}-\eta_{i,i+1}\mathbf u_{i}\right]^2. 
\label{ElasticGauge}
\end{equation}
The invariance of $(\epsilon_i u_i)$ under orientation transformation  translates into a $\mathbb{Z}_2$ 
gauge symmetry of the elastic energy: 
$\lbrace \eta_{i,i\pm1}\to-\eta_{i,i\pm1},\,\theta^{\rm S}_i\to-\theta^{\rm S}_i,\,
\theta^{\rm T}_i\to\theta^{\rm T}_i\rbrace$. 
 Following the procedure sketched in Fig.~\ref{Fig2}, Eq.~\eqref{ElasticGauge} can be
simplified by gauging away the $\eta_{i,i+1}$ at all sites but one, at $i\equiv i_{\rm G}$
where $\eta_{i_{\rm G},i_{\rm G}+1}=\mathcal O$.
For this gauge choice, 
$\mathcal E$ takes the compact form: 
${\mathcal E}=
    \frac{K}{2a}\sum_{i} ( \left[\theta^{\rm S}_{i+1}-\theta^{\rm S}_{i}\right]^2
                        + \left[\theta^{\rm T}_{i+1}-\theta^{\rm T}_{i}\right]^2 )
  + \frac{K}{a} \left(1-{\mathcal O}\right)\theta^{\rm S}_{i_{\rm G}}
                \theta_{i_{\rm G}+1}^{\rm S}$, 
where we have implicitly assumed $w/L$ to be vanishingly small and left  finite-size
geometrical corrections to future work~\cite{Kamien_Review}. 
The last term of this expression accounts for the coupling between the topological  charge $\mathcal O$ and the shear angle at the unspecified site $i_{\rm G}$. 
Continuum elasticity  then follows  from the  limit $a\to 0$ in Eq.~\eqref{ElasticGauge}:
\begin{multline}
{\mathcal E}(\lbrace\theta^{\rm S},\theta^{\rm T}\rbrace)=\frac{K}{2}\int_0^L \! \!  \left(\partial_s \theta^{\rm S}\right)^2+ \left(\partial_s \theta^{\rm T}\right)^2 {\rm d}s\\
+ (1-\mathcal O)\lim_{a \to 0} \frac{K}{a}
\left[(\theta^{\rm S})^2+a\theta^{\rm S}\partial_s\theta^{\rm S}
\right]_{s=s_{\rm G}}.
\label{Eq:Celasticity}
\end{multline}
For orientable strips, one recovers the familiar harmonic energy of elastic bodies. In contrast, for M\"obius strips where ${\mathcal O}=-1$, the  topological term in Eq.~\eqref{Eq:Celasticity} constrains the continuous shear deformations to vanish at $s=s_{\rm G}$ 
\footnote{Note that an alternative description consists in allowing $\theta^{\rm S}$ to be discontinuous. 
 The topological term would then imposes $\theta^{\rm S}(s_{\rm G})=-\theta^{\rm S}(s_{\rm G}+L)$:
 shear deformations would be antiperiodic functions of $s$.}.  Two comments are in order:  Firstly, unlike shear deformations, we find that twist deformations are insensitive to orientability and obey uncronstrained harmonic elasticity. Secondly, we stress that the location of the zero-shear point $s_{\rm G}$ is an independent and crucial gauge degree of freedom that must be dealt with when computing the fluctuations and  mechanical response of  non-orientable elastic ribbons as illustrated below. 
%
\begin{figure}
\includegraphics[height=7cm,angle=0]{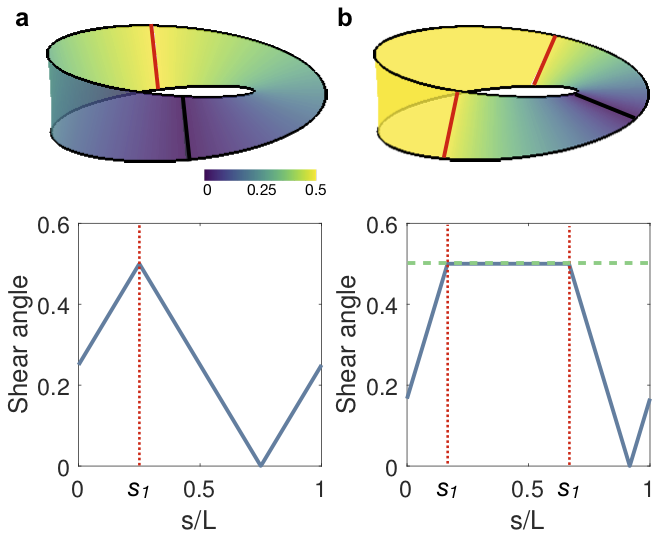}
\caption{{\bf Nonlinear  response to shear stress.} 
Top panels: the color indicates the magnitude of the shear angle along the M\"obius strips. The red lines show the positions of the applied stresses, and the dark line the position of $s_{\rm G}$. Bottom panels: corresponding plots of $\theta^{\rm S}(s)$. 
{\bf a.}  Response to a point-wise stress source $\sigma(s)=\sigma\delta(s-1/4)$ with $\sigma=K$. The shear angle decays linearly from the stress source and vanishes at $s=3/4$.  
{\bf b.}  Nonlinearity: Response to two point-wise stress sources $\sigma^{\rm S}=K[\delta(s-1/6)-\delta(s-(1/6+1/2))]$. The  deformations computed from the minimization of the elastic energy as explained in Appendix~\ref{NStressSources}  vanish at $s_{\rm G}= {11}/{12}$. The deformations computed from the minimization of the elastic energy (solid line)  are markedly different from the linear supperposition of two responses to two individual point sources (dashed line).
}
\label{Fig3}
\end{figure}
\begin{figure}
\includegraphics[height=6.95cm,angle=0]{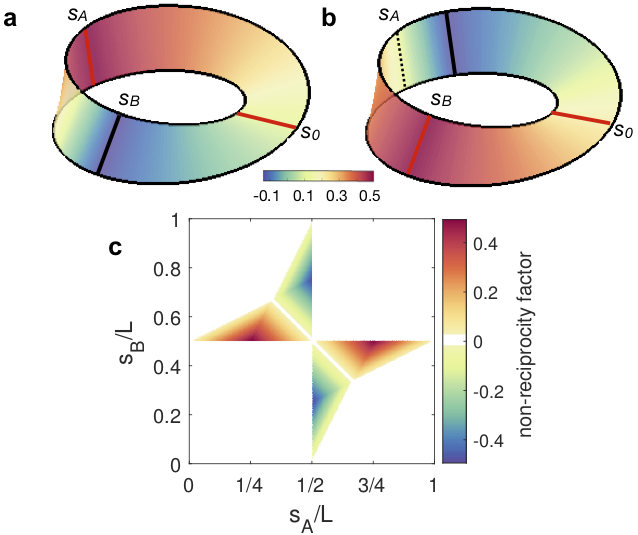}
\caption{{\bf Nonreciprocal response to shear stress.} 
A M\"obius strip is sheared by a localized stress distribution $\sigma_0=\sigma\delta(s-s_0)$ with $\sigma=K$ and $s_0=0$. {\bf a.}  Overall excess shear deformations $\theta_A(s)$ due to an additional stress  $\sigma_{A}=\sigma\delta(s-s_A)$   applied at $s_A=7/16$.  {\bf b.} Plot of the excess shear  deformations $\theta_B(s)$ measured when $\sigma_A$ is released and  $\sigma_{B}=\sigma\delta(s-s_B)$  is applied at ($s_B=23/32$).   Same color map as in {\bf a}. Together the plots reveal that $\theta_{A}(s_B)\neq\theta_B(s_A)$. {\bf c} Variations  of the non-reciprocity factor $\Delta\theta$ as a function of the locations of  $s_A$ and $s_B$.  The finite value of $\Delta\theta$ over a finite region of space proves that the Maxwell-Betti theorem breaks down on non-orientable elastic surfaces: non-oreinatble elasticity is non-reciprocal.
}
\label{Fig4}
\end{figure}
\begin{figure*}
\includegraphics[width=0.8\textwidth,angle=0]{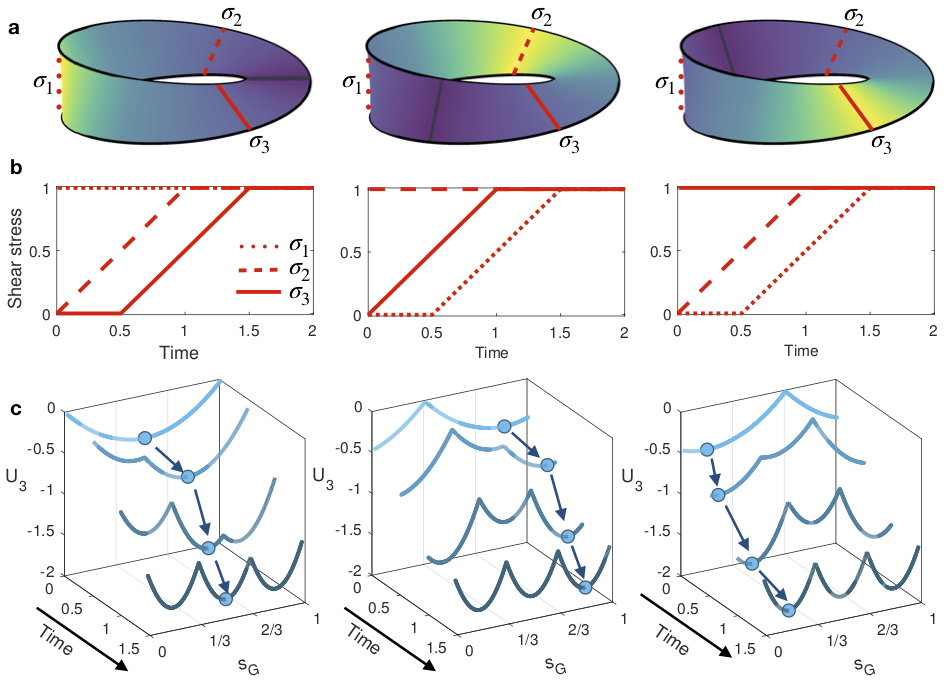}
\caption{{\bf Elastic Memory.} {\bf a.} Three different shear deformations are compatible with mechanical equilibrium when point-wise shear stresses of equal strength $K$ are applied at $s=1/6,\,1/2,$ and $5/6$. The colors indicate the magnitude of the shear deformations $\theta^{\rm S}$, same colormap as in Fig.~\ref{Fig3}. {\bf b.} Corresponding time variations of the stress amplitude. {\bf c} Spatial and temporal variations of the effective potential $U_3(s,t)$. The blue circle indicates the location of the instantaneous minimum of $U_3$ where $s_{\rm G}$ is trapped. }
\label{Fig5}
\end{figure*}
\subsection{Non-additive  elasticity}
\label{Section:Shear}
From now on the ribbon elasticity is prescribed by Eq.~\eqref{Eq:Celasticity}, and the constraint $\theta^{\rm S}(s_{\rm G})=0$. It then readily follows that   non-orientable ribbons cannot support any homogeneous shear deformation as anticipated in the introduction of Section~\ref{sectionII} and further discussed in Appendix~\ref{Appendixnohomogeneousstress}. 
The simplest mechanical stress we can consider is  a pointwise  shear  localized at an arbitrary position $s_1$:  $\sigma^{\rm S}(s)=\sigma\delta(s-s_1)$. 
The resulting deformations shown in Fig.~\ref{Fig3}a are computed minimizing $\mathcal E+\mathcal W$ with respect to {both} the shear and  gauge  degrees of freedom, where 
$\mathcal W=-\int_0^L \sigma^{\rm S}(s)\theta^{\rm S}(s) {\rm d}s$ is the work performed by the external stress, 
see Appendix~\ref{Appendix:reciprocity}. We  find a positive elastic response that vanishes at a single point located at  maximal distance from the stress source: 
\begin{align}
\theta_1(s;s_1)&=\frac{\sigma}{2K}\left|s-s_1-\sign(s-s_1)\frac{L}{2}\right| . 
\label{Eq:green}
\end{align}
 This simple expression  has a deep  consequence: the response of M\"obius strips to shear stresses is intrinsically nonlinear, although the local stress-strain relation is linear. 
 We establish this counter intuitive property by considering the case of two identical   stress sources: $\sigma^{\rm S}(s)=\sigma\left[\delta(s-s_1)+\delta(s-s_2)\right]$. 
 The  linear superposition of two $\theta_1$ functions  would result in strictly positive shear deformations  over the whole strip which is topologically prohibited as
 $\theta^{\rm S}$ must vanish at least at one point $s_{\rm G}$.  
 We therefore conclude that the response of M\"obius strips to shear stresses is not pairwise additive and therefore nonlinear.  This property is  illustrated in 
 Fig.~\ref{Fig3}b  where we compare the  shear angle $\theta_2(s;s_1,s_2)$ computed from the minimization of $\mathcal E+\mathcal W$ with respect to $\theta^{\rm S}$ and 
 $s_{\rm G}$ to that that derived from a mere superposition principle, see also Appendix~\ref{NStressSources}.
 
 We explain below the  practical consequence of this topological frustration.
 \subsection{Non-reciprocal   elasticity}
The static response of elastic bodies is generically reciprocal. In virtue of the so-called Maxwell-Betti theorem,  the deformations measured at a point $B$, as a result of a force  applied at a point $A$, are identical to the deformations measured at  point $A$ as a result of the same force  when applied at point $B$~\cite{Maxwell1864,Betti1872,Coulais2017}.   The mechanics of non-orientable surfaces, however, is not reciprocal. 
To prove this counterintuitive results, we consider  as a refence state a M\"obius strip sheared by a localized  source  $\sigma_0=\sigma\delta(s-s_0)$  causing a deformation $\theta_1(s;s_0)$. Let us now apply an additional stress  $\sigma$ at $s_A$, and measure the response at $s_B$: $\theta_A(s_B)=\theta_2(s_B;s_0,s_A)-\theta_1(s;s_0)$. 
We now release the stress applied at $s_A$,  apply as stress $\sigma$ at $s_B$, and measure the response at $s_A$: $\theta_B(s_A)=\theta_2(s;s_0,s_B)-\theta_1(s;s_0)$. The two corresponding excess shear angles are shown in Figs.~\ref{Fig4}a and~\ref{Fig4}b and are obviously different.  Following~\cite{Coulais2017}, we plot in Fig.~\ref{Fig4}c the non-reciprocity factor $\Delta\theta=[\theta_A(s_B)-\theta_A(s_B)]K/\sigma$ as a function of the locations of the two applied stresses ($s_A$ and $s_B$). We find that $\Delta\theta$ is  finite over a large fraction of the parameter space and extremal when the  stress sources are distant from $L/4$ and $L/2$ from $s_0$: the mechanical response of the strip is non reciprocal.
   {\color{bleu}  Two comments are in order. By contrast with the polar metamaterials considered in~\cite{Coulais2017}, here non-reciprocity does not rely on non-proportional response. The constitutive relation between stress and strain is linear, the strip is not unstable, and no floppy mode is excited. Non-reciprocity solely  stems  from the non-additive response of non-orientable strips. We also stress that non-reciprocity does not require  any fine-tuning of the strip geometry, or of the applied stresses: M\"obius strip mechanics is generically non reciprocal. } 
%
\subsection{Elastic memory}
In addition to be non-linear and, non-reciprocal, non-orientable elasticity is   multistable. This remarkable feature is demonstrated in Fig.~\ref{Fig5}a showing  three equilibrium shear deformations of a strip stressed by the same shear distribution: $\sigma(s)=\sigma_1 \delta(s-s1)+\sigma_2 \delta(s-s2)+\sigma_3 \delta(s-s3)$, with $\sigma_1=\sigma_2=\sigma_3=\sigma$.  
The difference between the three equilibrium states solely lies in the order used to switch on the three stresses, as illustrated~Fig.~\ref{Fig5}b.  The very origin of this elastic multistability stems from the trapping of $s_G$ at different locations between the $s_i$ points.
Integrating out the shear degrees of freedom, we derive in Appendix~\ref{NStressSources} the effective potential $U_3(s_G;\sigma_i(t))$ acting on  the zero-shear point $s_G$. We find that $U_3$ possesses as many minima as applied stress sources. Three possible shear deformations are therefore compatible with mechanical equilibrium, Fig~\ref{Fig5}c. We show in Fig~\ref{Fig5}c how the sequential increase of the three stresses selects one of the three minima and therefore the final mechanical state of the  M\"obius strip.
Non-orientable surfaces  offer a paradigmatic example of static mechanical memory. Information is coded and stored by the temporal variations of the stress. Information is read measuring the shear angle, and deleted releasing the applied stresses. 

\begin{figure*}
\includegraphics[width=0.9\textwidth]{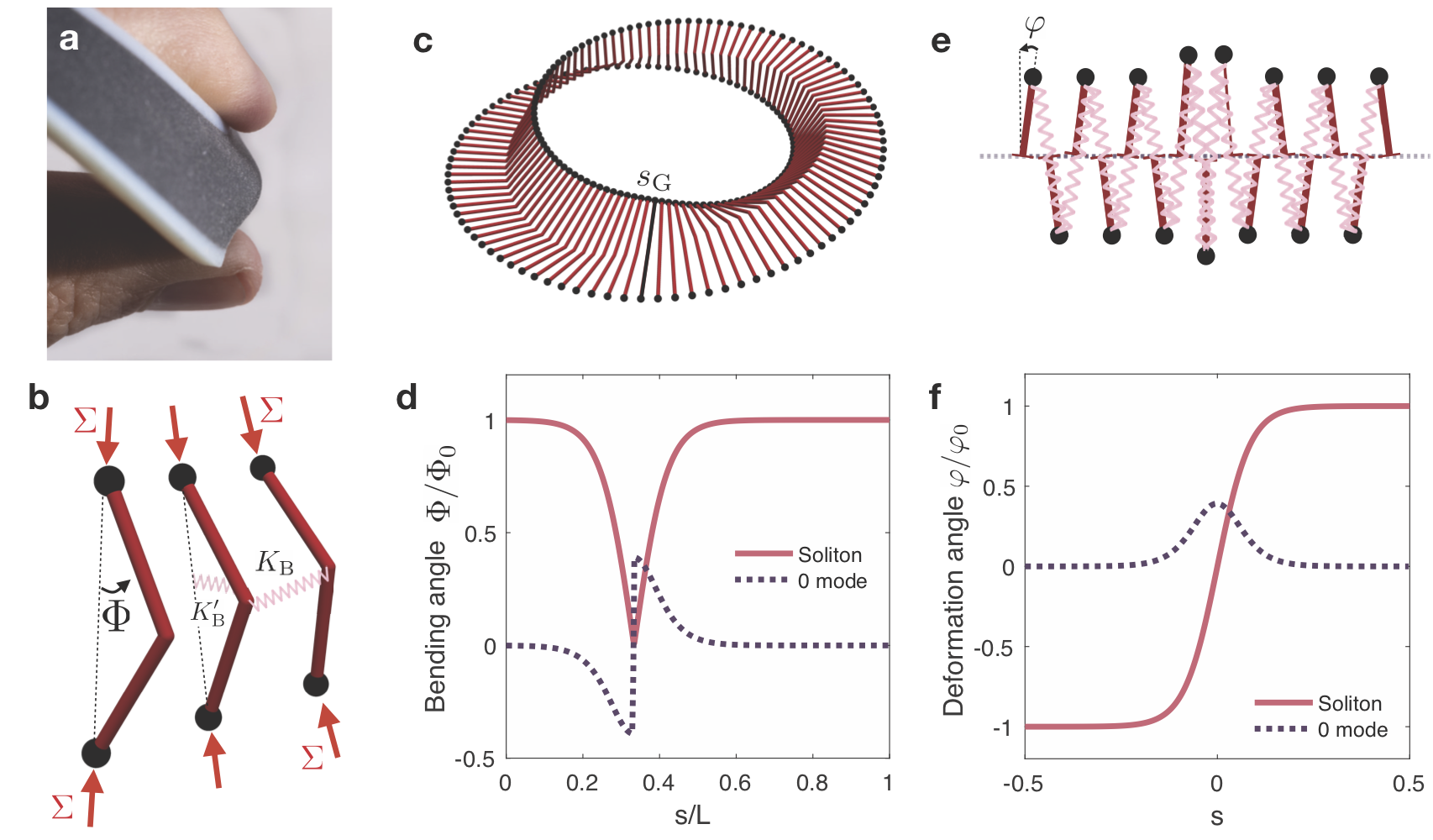}
\caption{{\bf From buckled a M\"obius strip to topological interface states.} 
{\bf a.} A 3D printed  M\"obius strip buckles upon application of a local compressive stress. {\bf b.} Sketch of the simplified buckling-elasticity model. A M\"obius ladder is composed of rigid bars of length $\ell$  connected by  soft hinge of stiffness $K'_{\rm B}$. The hinges are themselves coupled by harmonic springs of stiffness $K_{\rm B}$. The deformation of the $i^{\rm th}$ hinge is 
 parametrized by the angle $\Phi_i$. $\Sigma$ indicates the strength of the applied compressive stress.
  {\bf c.} Heterogeneous buckling pattern of a M\"obius ladder. {\color{bleu}The location of the undeformed bar at $s=s_{\rm G}$ is indicated with a dark color}. $\Phi_0=\pi/6$, $\xi=L/5$.
  {\bf d.} Solid line: Shape of the soliton, i.e. variations of the bending angle $\Phi_{\rm B}(s)$. $s_{\rm G}$ topologically protect the strip from homogeneous bending.  Dashed line: Shape of the floppy mode bound to the solitonic excitation. $\Phi_0=1$, $\xi=0.12L$.
    {\bf e.} Sketch of the mechanical SSH model introduced in~\cite{KaneLubensky}. This simple mechanicla metamaterial is composed of canted rigid bar free to rotate around a single axis are connected by harmonic springs, see~\cite{Chen2014} for an experimental realization.
    {\bf f} Solid line, the interface between two topologically distinct mechanical metamaterials is a $\phi^4$ kink. Dashed line: topological interface mode of the SSH model. $\varphi_0=1$, $\xi=0.12$. 
 }
\label{Fig6}
\end{figure*}

\section{Buckling a M\"obius strip}
\subsection{Solitary buckling waves}

We now show how heterogeneous deformations emerge from homogeneous stresses.  
To do so, we address the consequences of non-orientability on the bending deformations of  M\"obius strips, see Fig.~\ref{Fig6}a. 
We consider a simplified description where the strip is modelled by a ladder made of flexible hinges of length $\ell$ as
sketched in Fig.~\ref{Fig6}b. {\color{bleu} For sake of simplicity, we restrain ourselves to bending deformations along the normal vector which   naturally couple to the ribbon orientation.}   
The total elastic energy ${\mathcal E}_{\rm B}$ is composed of three terms: 
(i) the conformation the $i^{\rm th}$ hinge is defined by the angle $\Phi_i$ and is associated with a harmonic bending energy $\sim \frac{1}{2a} K_{\rm B}'\Phi_i^2$,  
(ii) a harmonic coupling between the hinges adds a contribution $\sim \frac{1}{2a} K_{\rm B}(\Phi_{i+1}-\eta_{i,i+1}\Phi_{i})^2$, and 
(iii) applying an external compression  load $\Sigma_i$ contributes {\color{bleu}to a mechanical work defined as the scalar product between the applied force and the resulting displacement:} $(\Sigma_i a\ell)(\cos\Phi_i-1)$.  
We are now equipped to tackle the classical Euler buckling problem: the bending instability of a M\"obius strip in response to a homogeneous compression. 
We first construct a continuum description of ${\mathcal E}_{\rm B}$ 
following the same procedure as in Section~\ref{Section:shear}, gauging away the $\eta_{i,i+1}$ variables, taking the continuum limit and restraining ourselves to  deformations close to the onset of buckling. ${\mathcal E}_{\rm B}(\lbrace \Phi \rbrace )$ then takes the compact form:  
\begin{multline}
{\mathcal E}_{\rm B}=\int\!\frac{K_{\rm B}}{2}\left(\partial_s \Phi\right)^2
+U^2(\Phi)\, {\rm d}s\\
+ (1-\mathcal O) \lim_{a\to 0} \frac{K_{\rm B}}{a}\left(\Phi+a\Phi\partial_s\Phi\right)^2_{s=s_{\rm G}},
\label{Eq:phi4}
\end{multline}
where the quartic potential  
\begin{equation}
U(\Phi)=\frac{\sqrt{K_{\rm B}}}{\xi}(\Phi^2-\Phi_0^2)
\end{equation}
 is classically parametrized by the scale 
$\xi^{2}=24a^2(K_{\rm B}/K'_{\rm B})(\Sigma_0/\Sigma)$  over which bending deformations occur, 
and the distance to the critical buckling load of an isolated hinge: $\Phi_0^2=6(1-\Sigma_{\rm 0}/\Sigma)$, with $\Sigma_{\rm 0}a^2\equiv K'_{\rm B}/\ell$.  
The last term of Eq.~\eqref{Eq:phi4} is the gauge fixing term which constrains $\Phi(s)$  to vanish at a point $s_{\rm G}$, thereby  protecting  M\"obius strips from homogeneous buckling. 
However, unlike shear stresses, the compression $\Sigma$ can be applied uniformly along the ribbon, and does not break translational invariance. $s_{\rm G}$ is then a free degree of freedom that parametrizes the broken-symmetry deformations. 

The buckling  patterns minimize  ${\mathcal E}_{\rm B}$
with the constraint $\Phi(s_{\rm G})=0$.
This minimization is performed  using a dynamical-system analogy elaborated in Appendix~\ref{Appendix:A}. 
In short, the strip remains flat until $\Sigma$ exceeds $\Sigma_{\rm c}=\Sigma_0\left[1+(K_{\rm B}/K'_{\rm B})\pi^2(a/L)^2\right]$. Above $\Sigma_{\rm c}$ it undergoes a  buckling transition and deforms into  the inhomogeneous pattern illustrated in Fig.~\ref{Fig6}c. 
The bending angle $\Phi$ remains close to $\Phi_0$  
everywhere except in a region of size $\xi/\Phi_0$ around  $s_{\rm G}$ where it vanishes.   
The buckling pattern is the norm of a $\Phi^4$ kink centered  on $s_{\rm G}$~\cite{ChaikinLubensky}.  In the
  limit of 
long  strips $\Phi_{\rm B}$ reduces to:
 \begin{equation}
     \Phi_{\rm B}(s-s_{\rm G})= \pm \Phi_0\left|\tanh\left(\frac{\sqrt{2}\Phi_0}{\xi}(s-s_{\rm G})\right)\right|+{\mathcal O}\left(\frac{\xi}{L}\right).
     \label{Eq:kink}
 \end{equation}
where the  sign of the solution reflects the arbitrary choice of orientation of the ribbon. 
The exact solution beyond the very long strip approximation does not bring more insight
and is left to Appendix~\ref{Appendix:A}. 

Remarkably, both the $\mathbb{Z}_2$ gauge symmetry and translational invariance are spontaneously broken at the onset of buckling. The ground state of $\mathcal E_{\rm B }$ is continuously degenerate leaving  the bending direction and  the location of the flat section $s_{\rm G}$ undetermined. As a consequence the buckling patterns  are free to translate around the strips. More quantitatively, having macroscopic systems in mind, we  now consider the inertial dynamics of the buckled strip 
 described by the continuum Hamiltonian:
 {\color{bleu}
 \begin{equation}
 {\mathcal H}_{\rm B}=\int\!\frac{I}{2}\left[(\partial_t\Phi)^2+\frac{K_{\rm B}}{2}(\partial_s\Phi)^2+U^2(\Phi)\right]\,{\rm d}s,
 \label{Eq:HB}
 \end{equation}
 }
where $I$ is the local moment of inertia.
As in the static case, $ {\mathcal H}_{\rm B}$ is complemented by the constraint $\Phi(s_{\rm G}(t),t)$=0. 
The existence of solitary waves readily follows from the Lorentz invariance of ${\mathcal H}_{\rm B}$. 
The solitary waves are deduced from Eq.\eqref{Eq:kink} by 
a Lorentz boost: 
$\Phi(s,t)=\Phi_{\rm B}(\gamma[s-vt])$, where 
$\gamma\equiv[1-(v/c)^2]^{-1/2}$~\cite{ChaikinLubensky}, 
and the propagation speed $v$ satisfies $v^2<c^2=K_{\rm B}/I$. 
The free propagation of these solitary bending waves restores   translational and gauge invariance of Eq.~\eqref{Eq:phi4}: 
moving the topologically protected section $s_{\rm G}$
 along the strip corresponds a mere gauge transformation which operates at zero energy cost. 
 We note that travelling kinks of the very same nature were first found theoretically and illustrated experimentally in soap films forming non-orientable minimal surfaces~ \cite{Pesci2016}. 
 In the context of topological mechanics, spectacular zero-energy mechanisms having a similar solitonic structure were also found at the interface between open one-dimensional isostatic lattice  having topologically distinct band spectra~\cite{Chen2014,Vitelli2014}. 
 In the next section, we show that the latter resemblance is the first hint of a deeper connexion between the topological mechanics of non-orientable ribbons and that of one-dimensional isostatic  metamaterials.%

\subsection{From buckled M\"obius strips to SSH topological  insulators}
We characterized above the orientability of the ribbon by the invariant $\mathcal O$, and showed  that $\mathcal O = -1$ implies the existence of solitary bending waves. 
Here, we show that these excitations are characterized by their own topological number $n$ which we relate to that of interface states between topological insulators. The standard topological characterization of both phononic and electronic excitation was established for lattice models and does not apply to continuous elasticity~\cite{Mao2018,Hasan2010}. We circumvent this technical obstacle following Ref.~\onlinecite{Vitelli2014}. Resorting to an index theorem applied to the  linearized elasticity of the strip, we establish that the soliton carries a topological charge  $n=1$ that counts the zero energy translational modes.

In practice, we introduce the linear bending fluctuations of the ribbon around a static buckled state \eqref{Eq:kink}: $\Phi(s,t)=\Phi_{\rm B}(s-s_{\rm G})+\Psi(s,t)$, and deduce the dynamics of $\Psi$ by linearizing Eq.~\eqref{Eq:HB}:
\begin{align}
\partial^2_t\Psi(s,t)=-{\mathbb D}\Psi(s,t),
\label{Eq:dynamics}
\end{align}
where ${\mathbb D}=\partial_s^2-2U'^2(\Phi_{\rm B})-2U(\Phi_{B})U''(\Phi_{\rm B})$.  The topological properties of mechanical vibrations are revealed by the ''square root''  of the dynamical operator $\mathbb D$~\cite{KaneLubensky}. In the limit of very long yet finite ribbons, the dynamical operator can be recast into the factorized form (see Appendix~\ref{Appendix:B}):
\begin{align}
\mathbb D={\mathbb Q}^{\dagger}{\mathbb Q}+{\mathcal O}\left[(\xi/L)^2\right],
\label{Eq:D}
\end{align}
where ${\mathbb Q}^\dagger=\partial_s+\textrm{sign}(s-s_{\rm G})\sqrt{2}U'(\Phi_{\rm B})$, 
and ${\mathbb Q}=-\partial_s+\textrm{sign}(s-s_{\rm G})\sqrt{2}U'(\Phi_{\rm B})$.
The soliton is then associated to  
the topological index $n$ that counts the zero modes of  $\mathbb D$ making a distinction between floppy modes and self-stress states~\cite{KaneLubensky,Mao2018,Vitelli2014}:
\begin{equation}
n={\rm dim\,ker}\,{\mathbb Q}-{\rm dim\,ker}\,{\mathbb Q}^{\dagger}.
\label{Eq:index}
\end{equation}
We compute $n$ by determining explicitly the kernel  of  the two linear operator ${\mathbb Q}$ and ${\mathbb Q}^\dagger$. In the limit $L/\xi\gg1$ the corresponding eigenequations  reduce to:
\begin{equation}
\partial_s \ln \Psi_{\pm}=\pm\partial_s \ln[\partial_s\Phi_{\rm B}(s)].
\label{Eq:logs}
\end{equation}
Solving Eq.~\eqref{Eq:logs} on the circle, we find that the kernel of ${\mathbb Q}^\dagger$ is trivial,
 while $\mathbb Q$ has a one-dimentional kernel:
 $\Psi_+=\delta s_{\rm G} \partial_s\Psi_{\rm G}$, where $\delta s_{\rm G}$ is a  parametrization factor.  This 
 solution corresponds to an infinitesimal translation of the soliton: 
 $\Psi_+(s;s_{\rm G})=\Phi_{\rm B}(s-s_{\rm G}-\delta s_{\rm G})-\Phi_{\rm B}(s-s_{\rm G})$ and is plotted in Fig.~\ref{Fig6}d.  We stress that this 
 translational mode is a floppy mode that operates, by definition, at zero energy cost. The  topological 
 index defined by Eq.~\eqref{Eq:index} being non trivial ($n=1$), it ascertains the topological nature of the floppy modes and of 
 the associated solitary wave. 
 
 {\color{bleu}This zero mode is reminiscent of the boundary  states predicted by topological band theory at the interface between materials having topologically inequivalent eigenspaces~\cite{Hasan2010,Bernevig,KaneLubensky}. These two types of zero modes are however essentially different. More precisely,  Eqs.~\eqref{Eq:dynamics} and \eqref{Eq:D} are similar to the  equations describing the vibrations of the interface between two SSH mechanical metamaterials illustrated in Fig.~\ref{Fig6}e, see~\cite{Chen2014,Vitelli2014}. In this different context, the existence of an interfacial zero mode is guaranteed by the imbalance between the number of self-stress and floppy modes given by the Kane-Lubensky generalization of the Maxwell-Caladine index~\cite{KaneLubensky}.  In the settings of Fig.~\ref{Fig6}e, the Kane-Lubensky count is equal to 1 thereby imposing the binding of a floppy mode to the interface. By contrast, the buckled M\"obius ladder sketched in Fig.~\ref{Fig6}c is a closed isostatic system with a vanishing Maxwell-Caladine index.  Therefore the existence of its zero mode  is not captured by the Kane-Lubensky-Maxwell-Caladine index which disregards the gauge degrees of freedom associated to orientability. Beyond the specifics of mechancial systems, this counterintuitive observation prompts us to reconsider to very concept of the bulk-boundary correspondence of topological band theory when applied to non-orientable (meta)materials~\cite{Beugeling2014,Ryu,Witten}.
 }

\section{Conclusion and Perspectives}
We have demonstrated how to surpass  the native  properties of  materials without resorting to geometrical tuning. Constructing  a minimal elastic theory for M\"obius strips, we have established that  non-orientability makes their  local mechanics  non-linear,  non-reciprocal and  capable of memorizing its stress history. Investigating their simplest bending instability, we have demonstrated how non-orientability guarantees the existence  of a topological phase that supports zero-enery solitons. {\color{bleu}This mechanical phase, without known condensed-mater counterparts,  begs for a generalization of the current bulk-boundary correspondance in topological materials~\cite{Kitaev2009,Ryu2010,Chiu2016,Zhang2019,Tang2018,Vergniory2018}.}

Our  main predictions are elaborated building on prototypical models,  we therefore expect their experimental implications to extend beyond the specifics  of mechanical systems. In particular, the relation between nematic elasticity and ${\mathbb Z_2}$ gauge theories was realized in the early 90's by Lammert {\it et al.} in the context of phase ordering, but to the best of our knowledge has remained virtually uncharted~\cite{Lammert1995}. We stress here that our central equation Eq.~\ref{Eq:Celasticity} also describes the Frank energy of  non-orientable nematic films, and  can be generalized to describe  nematic elasticity around a disclination~\cite{Alexander2012}. A remarkable experimental realization of a non-orientable nematic liquid crystal was provided by self-assembled viral membranes where rod-like units self-organize into M\"obius conformations at the membrane edge~\cite{Gibaud2017}. Beyond elasticity, we also envision our prediction to be relevant to M\"obius configurations of light polarization~\cite{Bauer2015,Bliokh2019}, and to transport in twisted  nano crystals~\cite{Tanda2002}. 
\section*{acknowldegments}
This work was supported by IdexLyon breakthrough program and ANR WTF grant. 
We thank Michel Fruchart and Krzysztof Gawedzki for stimulating discussions, William Irvine and Jeffrey Gustafson for help with the fabrications  of M\"obius strips, and Corentin Coulais for introducing us to nonreciprocal mechanics and for insightful suggestions. 
D.B. and D.C  have equally contributed to all aspects of the research.

\appendix
\section{Response to  shear }
\label{Appendix:reciprocity}
\subsection{No homogeneous shear stress}
\label{Appendixnohomogeneousstress}
We showed in Section~\ref{Section:Shear} that a M\"obius strip cannot support any homogeneous shear deformation. 
The situation is even more constrained as no uniform shear stress can be applied.  The shear stress $\sigma^{\rm S}$ and $\theta^{\rm S}$ are  conjugated variables, and the mechanical work associated to  shear is given by ${\mathcal W}=-\int\sigma^{\rm S}(s)\theta^{\rm S}(s)\,\rm ds$. As $\mathcal W$ must be independent on the arbitrary definition of the ribbon orientation, $\sigma^{\rm S}$ and $\theta^{\rm S}$ must obey the same transformation rules upon any change in the ribbon orientation: $\sigma(s)$ is also topologically constrained to vanish at $s_{\rm G}$.
\subsection{Response to a  point-wise shear stress}
We consider the response to the shear-stress distribution given by: $\sigma^{\rm S}=\sigma_1 \delta(s-s_1)$. For sake of clarity, units are here chosen so that $L=1$.  
The equilibrium configuration is obtained minimizing  the total energy
$\mathcal E+\mathcal W$ defined in Section~\ref{Section:Shear} with respect to both $\theta^{\rm S}$ and the gauge degree of freedom $s_{\rm G}$: 
\begin{equation}
\mathcal F=\mathcal E+\mathcal W=\frac{K}{2}\int_0^1  \left(\partial_s \theta^{\rm S}\right)^2
    -\int_0^1\sigma^{\rm S}(s)\theta^{\rm S}(s)\,\rm ds.
\label{Eq:EshearAppendix}
\end{equation}
We recall that $s_{\rm G}$ is the location of the strip section where $\theta^{\rm S}(s)$ is topologically constrained to vanish.  Within this framework, the two mechanical equilibrium conditions are: 
\begin{align}
\frac{\delta \mathcal F}{\delta\theta^{\rm S}} &=
-K\partial_s^2\theta^{\rm S}-\sigma^{\rm S}(s)=0,
\label{groundstate}
\\
\frac{\partial \mathcal F}{\partial s_{\rm G}} &= 0. 
\label{f1}
\end{align}
These equations are supplemented by the boundary conditions:
\begin{equation}
\theta^{\rm S}(s,s_1;s_{\rm G})=\theta^{\rm S}(s+1,s_1;s_{\rm G}),
\label{eq:App-boundary}
\end{equation}
and the topological constraint 
\begin{equation}
    \theta^{\rm S}(s=s_{\rm G},s_1;s_{\rm G})=0.
\label{eq:AppTopConstr}
\end{equation}

The algebra is simplified by redefining the origin of the curvilinear coordinate ($s\to \tilde s$) such that 
$\tilde{s}_{\rm G}=0$. The conditions (\ref{eq:App-boundary},\ref{eq:AppTopConstr}) then reduce to 
\begin{equation}
\theta^{\rm S}(\tilde{s}=0,\tilde{s}_1)=\theta^{\rm S}(\tilde{s}=1,\tilde{s}_1) . 
\label{eq:App-boundaryTopo}
\end{equation}
In this frame, the gauge degree of freedom then becomes the position of the applied stress 
$\tilde{s}_1 = s_1-s_{\rm G}$ (mod $1$).
Solving Eqs.~\ref{groundstate} and~\ref{eq:App-boundaryTopo}, we readily find that  
the shear deformations are given by 
$\theta_1^{\rm S}(\tilde{s};\tilde{s}_1)=  \sigma_1 G^{(1)}(\tilde{s};\tilde{s}_1)$ with 
\begin{equation}
G^{(1)}(\tilde{s};\tilde{s}_1) = 
-\frac{1}{2K}
\left(\left|\tilde{s}-\tilde{s}_1\right|+ (\tilde{s}_1 -1)\tilde{s} + (\tilde{s}-1)\tilde{s}_1 \right). 
\label{theta1}
\end{equation}
The corresponding total energy 
\begin{equation}
    \mathcal{F} = - \frac{\sigma_1^2}{2K} \tilde{s}_1 (1-\tilde{s}_1)
\end{equation} 
is minimized for $\tilde{s}_1 = \frac12$, {\it i.e.} for $s_{\rm G} = s_1 + \frac12 $ mod $1$. 
In other words, the point $s_{\rm G}$ where the shear deformation vanishes is maximally separated from the applied stress. Going back to the original frame, the static shear deformations at mechanical equilibrium 
are easily recast into: 
\begin{equation}
\theta_1(s,s_1)
=\frac{\sigma_1}{2K}\left|s-s_1-\frac{1}{2} \sign(s-s_1)\right|, 
\end{equation}
which corresponds to  Eq.~\ref{Eq:green} in the main text.

\subsection{Response to  $N$ localized shear sources}
\label{NStressSources}
We now consider the superposition of $N$ 
{ fixed} point-wise sources : $\sigma^{\rm S}(s)=\sum_{i=1}^N\sigma_i\delta(s-s_i)$. 
The equilibrium conformation $\theta^{\rm S}_N(s,\lbrace s_i\rbrace ; s_{\rm G})$ of the M\"obius 
strip satisfies the condition (\ref{groundstate},\ref{f1}) with the boundary conditions (\ref{eq:App-boundary},\ref{eq:AppTopConstr}). 
Working in the frame where $s_{\rm G}=0$, the solution of this equation is 
\begin{equation}
\theta_N^{\rm S}(\tilde{s};\lbrace \tilde{s}_i\rbrace)=\sum_{i=1}^{N}\sigma_i ~G^{(1)}(\tilde{s};\tilde{s}_i), 
\label{thetaN}
\end{equation}
where $\tilde{s}_i = s_i-s_{\rm G}$ mod $(1)$, {\it i.e.}
$\tilde{s}_i = s_i - s_{\rm G} + \Theta (s_{\rm G} - s_i)$ where 
$\Theta(s)$ is the Heaviside step function. 
At first sight Eq.~\ref{thetaN} resembles the mere superposition of independent Green functions and suggests  a typical linear response behavior.
However we have to keep in mind that the position $s_{\rm G}$ is yet to be determined to prescribe the  equilibrium deformations.
As a shift in the position $s_{\rm G}$ corresponds to a uniform translation of all the stress sources.  
we need to compute the equilibrium value of $\tilde{s}_1$, keeping all distances  $\tilde{s}_i - \tilde{s}_1$ fixed.  
Inspired by the classical calculation of the elastic interactions between inclusions in soft membranes and liquid interfaces (see e.g.~\cite{Bartolo2003}), 
we integrate over the shear degrees of freedom and derive the effective  potential $U_N(s_{\rm G})$ that controls the position of $s_{\rm G}$ along the strip. To compute $U_N$, it is convenient to 
solve a seemingly more complex problem where the strip  undergoes thermal fluctuations. The thermal statistics is then defined by the partition function 
\begin{multline}
\mathcal Z[\{\sigma_i \}]=
\int_0^1 {\rm d} s_{\rm G} \int\!{\mathcal D }\theta^{\rm S}
\\
e^{-\beta\left[\int\!{\rm d}\tilde{s}\,\frac{1}{2}(\partial_{\tilde{s}}\theta^{\rm S})^2
-\sum_i \sigma_i\theta^{\rm S}(\tilde{s}_i)\right ]},
\label{Z}
\end{multline}
where $\beta^{-1}\equiv K_{\rm B}T$ and the field $\theta^{\rm S}(\tilde{s})$ satisfies the condition 
\eqref{eq:App-boundaryTopo}. 
Integrating out the $\theta^{\rm s}$ degrees of freedom defines the effective potential $U_N(s_{\rm G})$:
\begin{align}
\mathcal Z & =\int_0^1\!{\rm d} s_{\rm G}\, e^{-\beta U_N(s_{\rm G})},
\\
\mathrm{with }\ \   U_N(s_{\rm G})& =-\frac 1 2\sum_{i,j}\sigma_i\sigma_j ~ G^{(1)}(\tilde{s}_i,\tilde{s}_j). 
\label{UN}
\end{align}
Going back to the original mechanics problem, i.e. taking the zero temperature limit in Eq.~\ref{UN}, we find the equilibrium position of $s_{\rm G}=0$ by minimizing $U_N(s_{\rm G})$. 
The non-linearity of the shear response of the M\"obius strip originates from this last minimization procedure, which translates the topological constraint.

\begin{figure}
\includegraphics[width=6cm,angle=0]{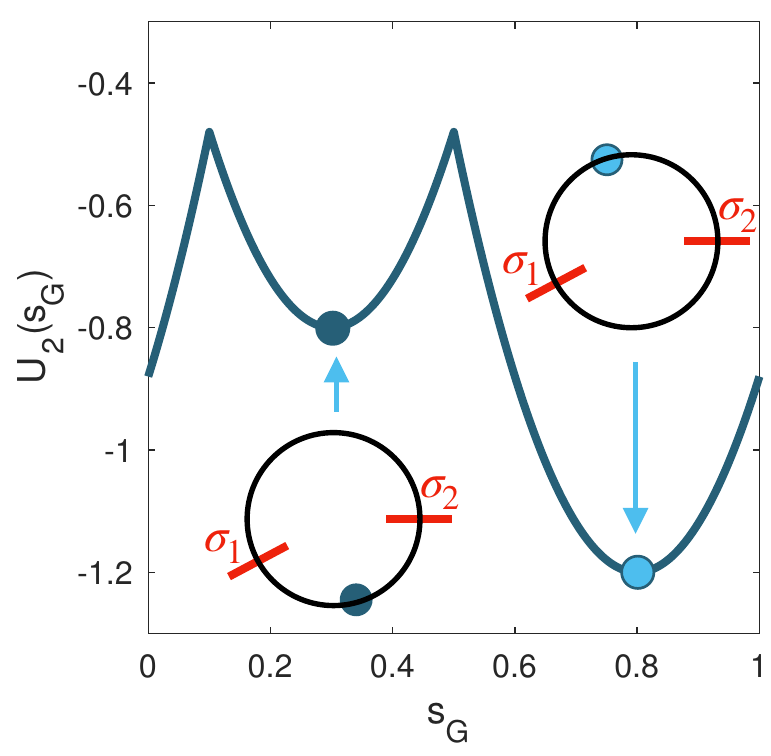}
\caption{{\bf Equilibrium positions of $s_{\rm G}$.} 
Effective potential $U_2(s_{\rm G})$  trapping the position of $s_{\rm G}$ along the strip when two identical shear stresses are applied at positions $s_1,s_2$. The blue circles indicate the two locally stable positions of  $s_{\rm G}$. $U_2$ is minimized when $s_{\rm G}$ is maximally distant from the appleid stresses. Illustration with $s_1=1/10$ and $s_2=1/2$ and $\sigma_1=\sigma_2=K$.
}
\label{FigAppendix}
\end{figure}

We  illustrate this method for two identical  stress sources located at $s_1$ and $s_2$ separated by a constant distance $\Delta$: $\sigma^{\rm s}(s)=\sigma\left[\delta (s-s_1)+\delta(s-s_2)\right]$, with $s_2=(s_1+\Delta)\,{\rm mod}(1)$ and find
\begin{multline}
U_2(s_{\rm G})=\\
\frac{\sigma^2}{2K}\left( |\tilde{s}_2-\tilde{s}_1|+(\tilde{s}_1+\tilde{s}_2)(\tilde{s}_1+\tilde{s}_2-2) \right)
\end{multline}
Minimizing $U_2(s_{\rm G})$,  we find two local minima satisfying $\partial_{s_{\rm G}}U_2=0$ 
at $s_{\rm G} = (s_1+s_2)/2$ mod $(1)$
and $s_{\rm G} = (s_1+s_2)/2 +\frac12$ mod $(1)$. 
They are sketched in Fig.~\ref{FigAppendix} and reflect the mirror symmetry of the problem.  
The  lowest energy conformation always corresponds to the value of $s_{\rm G}$ the further away from the stress sources. 
In the symmetric case, where $\Delta=\frac1 2$, the shear response possesses two degenerate 
equilibrium positions. 
With the knowledge of the position $s_{\rm G}$ the shear-deformation profile is fully determined. It is given by Eq.~\eqref{thetaN}, and illustrated in Fig.~\ref{Fig3} for various positions of $s_1,s_2$.

\section{Buckling patterns and solitary waves, a dynamical-system insight.}
\label{Appendix:A}
\begin{figure}
\includegraphics[width=6cm,angle=0]{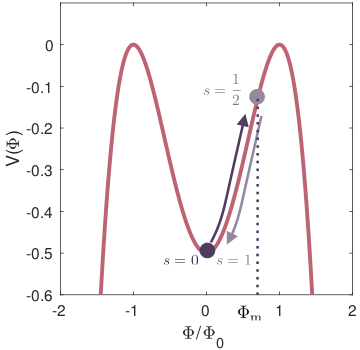}
\caption{{\bf Soliton shape: kicked particle in a potential.} 
The potential $V(\Phi)$ is plotted versus $\Phi$.  At time $s=0$ the particle is located in the potential well (dark circle). It is however not at rest as its velocity is non zero; it is  kicked uphill and reaches its maximal position at $s=\frac1 2$ (light circle). It then falls back to its initial position at $t=1$. The particle speed at $s=0$ and $s=1$ are opposite. $\Phi_0=1/2$, $\xi=1$. This feature translates into a slope discontinuity of the soliton shape at $s_{\rm G}$.
}
\label{FigAppendix2}
\end{figure}

We compute the shape  of buckled M\"obius strips making use of a dynamical system analogy. The expression of the elastic energy given by Eq.~\eqref{Eq:phi4} is indeed analogous to the Lagrangian of a classical particle of unit mass, and moving in a potential $V(\Phi)=-\xi^{-2}(\Phi^2-\Phi_0^2)^2$, where $\Phi$  indicates the particle position, $s$ the time, and $\partial_s \Phi$ the particle velocity, see Fig.~\ref{FigAppendix2}. Both the non-orientability constraint, and the finite size of the strip complexifies the dynamics of this seemingly simple dynamical system. We show below that the trajectories are not periodic and singular at $s_{\rm G}$.

 Without loss of generality we chose $s_{\rm G}=0$. Non-orientability therefore implies that $\Phi(0)=\Phi(L)=0$ regardless of the value of the particle speed $\partial_s\Phi(s_{\rm G})$. The trajectory $\Phi(s)$ is found noting that the mechanical energy $E_{\rm m}=\frac 1 2 [\partial_s\Phi(s)]^2+V[\Phi(s)]$ is a constant of motion. Noting $\Phi_{\rm m}=\max [\Phi(s)]$, the periodicity of the trajectory (reflecting the periodicity of the strip shape) imposes  $|\Phi_{\rm m}|<\Phi_0$. Otherwise, one would simultaneously have $\partial_s\Phi=0$ and $\Phi>\Phi_0$, thereby leading to runaway solutions.  Invariance upon time reversal of the particle Lagragian also imposes $\Phi_{\rm m}=\Phi(L/2)$. Therefore, the conservation of mechanical energy implies:
 \begin{equation}
\partial_s\Phi(s)=\pm{\sqrt{2[V(\Phi_{\rm m})-V(\Phi(s))]}}.
\label{Eq:dPhi}
 \end{equation}
 Let us consider solutions where $\Phi_{\rm m}>0$. The sign of $\partial_s\Phi(s)$ in Eq.~\eqref{Eq:dPhi} is then positive when $0<s<L/2$ and negative when $L/2<s<L$, and the inverse function $s=\Phi^{-1}[\Phi(s)]$ is  readily found integrating Eq.~\eqref{Eq:dPhi} on the two separate intervals:
 \begin{align}
s&=\pm\xi\int_0^{\Phi}\frac{{\rm d}x}{\sqrt{2(x^2-\Phi_{\rm m}^2)(x^2+\Phi_{\rm m}^2-2\Phi_0^2)}}\nonumber,\\
&=\xi\frac{1}{\sqrt{4\Phi_0^2-2\Phi_{\rm m}^2}}F[\arcsin(\Phi/\Phi_{\rm m},k)],\label{Eq:solutionPhi}
 \end{align}
 where $k\equiv\Phi_{\rm m}^2/(2\Phi_0^2-\Phi_{\rm m}^2)$ and $F(x,k)$ is the incomplete elliptic integral of the first kind. The final form of the trajectory follows from the definition $\Phi_{\rm m}\equiv\Phi(1/2)$ which imposes $\Phi_{\rm m}^2=2\Phi_0^2-2(\xi L^{-1})F(\pi/2,k)$.  We stress that  our gauge choice constraints $\phi(s)$ to vanish at $s_{\rm G}=0$ thereby imposing the derivative of $\Phi$ to be discontinuous at $s_{\rm G}$. The solution corresponds to two symmetric half $\Phi^4$ kinks defined on a compact interval  is plotted in Fig.~\ref{Fig6}d in the main text. One  last comments is in order. In the limit of large ribbons assembled from very stiff hinges, $\xi/L\ll1$, $\Phi_{\rm m}=\Phi_0$, and the integration of Eq.~\eqref{Eq:dPhi} results in the usual $\tanh$ profiles given by Eq.~\eqref{Eq:kink}. The buckling pattern corresponds to the symetrization of the usual $\Phi^4$ soliton.  
\section{Factorization of the dynamical operator.}
\label{Appendix:B}
We show how to factorize the dynamical operator ${\mathbb D}=\partial_s^2-2U'^2(\Phi_{\rm 0})-2U(\Phi_{\rm 0})U''(\Phi_{\rm 0})$ defined in Eq.~\eqref{Eq:dynamics}. As discussed above in Appendix~\ref{Appendix:A}, in the limit of infinitely long ribbons, $\Phi^\infty_{\rm m}=\Phi_0$ and $V(\Phi_{\rm m}^\infty)=E_{\rm m}=0$. The latter relation simplifies Eq.~\eqref{Eq:dPhi}:
\begin{equation}
\partial_s\Phi^\infty(s)=\textrm{sgn}(s-s_{\rm G})\sqrt 2 U[\Phi^\infty(s)].
\label{Eq:dsPhi}
\end{equation}
Together with the definition of $\mathbb D$, this relation implies the factorization $\mathbb D={\mathbb Q}^\dagger {\mathbb Q}$, with
\begin{align}
{\mathbb Q}^{\dagger}&=\partial_s+\textrm{sgn}(s-s_{\rm G})\sqrt{2}U'(\Phi^\infty_{\rm 0}),
\label{Eq:Qdagger}
\\
{\mathbb Q}&=-\partial_s+\textrm{sgn}(s-s_{\rm G})\sqrt{2}U'(\Phi^\infty_{\rm 0}),
\label{Eq:Q}
\end{align}
where  $\Phi^\infty_0$ is the shape of the unperturbed buckled ribbon. A $\xi/L$ expansion shows that this
form is  preserved for very long but {\em finite} ribbons.  This result is obtained  expressing
the ribbon shape as a linear perturbation of $\Phi^\infty$:
$\Phi_0=\Phi_{\rm 0}^\infty+(\xi/L)\tilde\Phi$. Evaluating $E_{\rm m}$, and keeping in
mind that $U(\Phi_0)=0$, we find: 
$E_{\rm m}=U^2(\Phi_0+\tilde\Phi_{\rm m})=U^2(\Phi_0)+2U(\Phi_0)U'(\Phi_0)\tilde{\Phi}_{\rm m}
    +{\mathcal O}[(\xi/L)^2]={\mathcal O}[(\xi/L)^2]$. 
The relations $E_{\rm m}=0$ and Eq.~\eqref{Eq:dsPhi} are hence preserved at first order in $\xi/L$. 
Therefore, even though the Hamiltonian $\mathcal H_{\rm B}$ defined in Eq.\eqref{Eq:HB}
does not enjoy the BPS symmetry of the continuum description of the isostatic chain of
linkages introduced in~\cite{Vitelli2014}, the corresponding dynamical matrix can still
be factorized as $\mathbb D={\mathbb Q}^\dagger {\mathbb Q}$, substituting $\Phi^{\infty}$
by $\Phi$ in Eqs.~\eqref{Eq:Qdagger} and Eqs.~\eqref{Eq:Q}.

%

\end{document}